\documentclass{article}

\usepackage{microtype}
\usepackage{graphicx}
\usepackage{subcaption}
\usepackage{booktabs} 

\usepackage{hyperref}

\usepackage{algpseudocode}

\usepackage[preprint]{icml2026}

\usepackage{amsmath}
\usepackage{amssymb}
\usepackage{mathtools}
\usepackage{amsthm}
\usepackage{ulem}

\usepackage[capitalize,noabbrev]{cleveref}

\theoremstyle{plain}

\theoremstyle{definition}

\theoremstyle{remark}

\usepackage[textsize=tiny]{todonotes}
\usepackage[table]{xcolor}

\usepackage{enumitem}

\newcommand{\cmark}{\text{\ding{51}}} 
\newcommand{\xmark}{\text{\ding{55}}} 
\usepackage{pifont} 

\newcommand{\jy}{}

\definecolor{IRBase}{RGB}{216,219,175}
\definecolor{VideoBase}{RGB}{231,220,197}
\definecolor{TSBase}{RGB}{236,205,189}
\definecolor{RefBase}{RGB}{220,200,213}
\definecolor{OursBase}{RGB}{176,194,211}
\colorlet{IRCol}{IRBase!40}
\colorlet{VideoCol}{VideoBase!40}
\colorlet{TSCol}{TSBase!40}
\colorlet{RefCol}{RefBase!40}
\colorlet{OursCol}{OursBase!40}

\newcommand{\err}[1]{\text{\scriptsize$\pm#1$}}

\icmltitlerunning{Operando Infrared Dynamics Generation from Static Spectra}

\begin{document}

\twocolumn[
  \icmltitle{From Static Spectra to \textit{Operando} Infrared Dynamics: Physics Informed Flow Modeling and a Benchmark}



  \icmlsetsymbol{equal}{*}

  \begin{icmlauthorlist}
    \icmlauthor{Shuquan Ye}{cuhk,slai}
    \icmlauthor{Ben Fei}{cuhk}
    \icmlauthor{Hongbin Xu}{mit}
    \icmlauthor{Jiaying Lin}{ust}
    \icmlauthor{Wanli Ouyang}{cuhk,slai}
  \end{icmlauthorlist}

  \icmlaffiliation{cuhk}{Department of Information Engineering, The Chinese University of Hong Kong}
  \icmlaffiliation{ust}{Department of Computer Science and Engineering, The Hong Kong University of Science and Technology}
  \icmlaffiliation{slai}{Shenzhen Loop Area Institute}
  \icmlaffiliation{mit}{Massachusetts Institute of Technology}
  \icmlcorrespondingauthor{Ben Fei}{benfei@cuhk.edu.hk}
  \icmlcorrespondingauthor{Hongbin Xu}{hbxu@mit.edu}

  \icmlkeywords{Machine Learning, ICML}

  \vskip 0.3in
]



\printAffiliationsAndNotice{}  

\begin{abstract}
  The Solid Electrolyte Interphase (SEI) is critical to the performance of lithium-ion batteries, yet its analysis via Operando Infrared (IR) spectroscopy remains experimentally complex and expensive, which limits its accessibility for standard research facilities.
  To overcome this bottleneck, we formulate a novel task, Operando IR Prediction, which aims to forecast the time-resolved evolution of spectral ``fingerprints'' from a single static spectrum. To facilitate this, we introduce OpIRSpec-7K, \textbf{the first} large-scale operando dataset comprising 7,118 high-quality samples across 10 distinct battery systems, alongside OpIRBench, a comprehensive evaluation benchmark with carefully designed protocols.
  Addressing the limitations of standard spectrum, video, and sequence models in capturing voltage-driven chemical dynamics and complex composition, we propose Aligned Bi-stream Chemical Constraint (ABCC), an end-to-end physics-aware framework. It reformulates MeanFlow and introduces a novel Chemical Flow to explicitly model reaction trajectories, employs a two-stream disentanglement mechanism for solvent-SEI separation, and enforces physics and spectrum constraints such as mass conservation and peak shifts. ABCC significantly outperforms state-of-the-art static, sequential, and generative baselines. ABCC even generalizes to unseen systems and enables interpretable downstream recovery of SEI formation pathways, supporting AI-driven electrochemical discovery.
\end{abstract}

\section{Introduction}

The global transition toward sustainable energy and electromobility has positioned lithium-ion batteries (LIBs) as a cornerstone of modern technology.
Central to battery performance, longevity, and safety is the Solid Electrolyte Interphase (SEI), a complex, dynamic and unstable passivation layer formed by the reductive decomposition of electrolytes. 
Despite decades of research, the SEI remains a ``black box'' due to its transient, multi-component nature and its sensitivity to the local electrochemical environment.

\begin{figure}[t]
    \centering
    \includegraphics[width=0.98\linewidth]{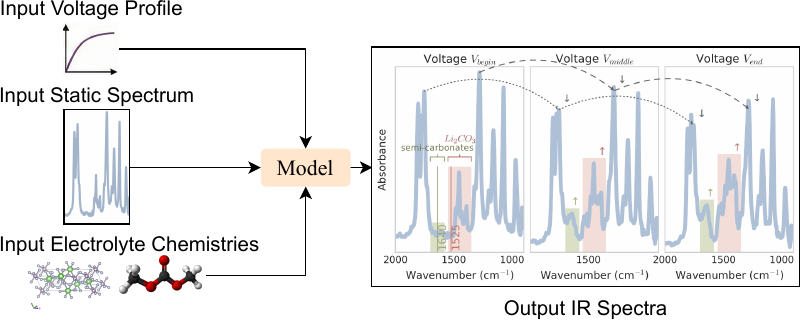} 
    \caption{\jy{We formulate a novel task: Operando IR Prediction. Starting from only a single, easily obtainable static spectrum with specific voltage profiles and electrolyte chemistries, the goal is to predict the evolution of IR spectra.} This output absorbance spectra is from Cu-PVDF//LP30//Li metal cell, where LP30 is 1\,M LiPF$_6$ in EC:DMC (1:1\,v/v), shown at three representative voltages ($V_{\text{begin}}$, $V_{\text{middle}}$, $V_{\text{end}}$). Arrows and trajectories provide a visual guide to spectral dynamics with different voltages, $\uparrow$ for formation and $\downarrow$ for decomposition. The highlighted windows isolate the $\nu$C=O region, \textit{e.g.}, semi-carbonates (LEDC, LMC, etc.) (green rectangle) and Li$_2$CO$_3$ absorb (orange rectangles). }
    \label{fig:teaser}
\vspace{-0.3cm}
\end{figure}

To decode the dynamic of SEI, Operando Infrared (IR) spectroscopy has emerged as the gold-standard diagnostic tool, offering direct observations into the dynamic chemical transformation at the electrode-electrolyte interface.
Unlike traditional ex-situ methods that need cell disassembly, which destroys the SEI's native structural integrity, Operando IR captures the real-time evolution of molecular vibrations under actual electrochemical environments.
It allows the mapping of specific IR absorption to the formation of organic and inorganic species, providing a high-fidelity chemical ``fingerprint'' of the interface as it evolves.

Despite its power, the adoption of Operando IR is bottlenecked by the extreme technical demands and expense~\cite{Leau2025Tracking}, especially for liquid-solid interfaces in LIBs~\cite{Li2020Operando}, leaving it accessible to only a handful of elite laboratories worldwide~\cite{AMARAL2023472}.

While recent machine learning advances have shown promise in predicting static spectra~\cite{ABDULAL2024141603,bhatia2025mace4irfoundationmodelmolecular,ye2020machine,mcgill2021predicting}, the community lacks the tools to model the dynamic evolution of electrochemical interfaces.
To make Operando IR more accessible, we first formulate and investigate a novel and urgent task: Operando IR Prediction. 
Starting from only a single, easily obtainable static spectrum, the goal is to forward-predict the evolution of IR spectra, driven by specific voltage profiles and electrolyte chemistries.

To address the scarcity of operando experimental data, we introduce the first large-scale Operando Spectrum Dataset, \textbf{OpIRSpec-7K}, featuring 7,118 high-quality samples across 10 distinct battery systems. 
We then establish a comprehensive benchmark, \textbf{OpIRBench}, using two rigorous settings: random splitting to evaluate sequence prediction accuracy, and system splitting to challenge the model’s generalization across unseen novel chemical compositions. 
We formulate our evaluation protocol to transcend standard numerical errors and similarity metrics by incorporating a dynamic distortion factor, which is specifically to capture abrupt phase mutations that are common in SEI evolution. To facilitate the use of state-of-the-art generative models, we propose an efficient spectrum-waveform auto-conversion pipeline. This allows mapping between one-dimensional spectral signals and two-dimensional energy waveform figures, enabling the seamless application of powerful video generation to the task. This multifaceted benchmark provides the first standardized foundation for AI-driven discovery in electrochemical interphase dynamics.

While standard IR prediction is mature, it is static and does not directly extend to this forward-predict operando spectral dynamics. Generic sequence and video models also miss spectroscopy specific physics. 
This gap arises for three reasons: \textbf{(1)} Operando spectra exhibit voltage-driven chemical dynamics; signals evolve as a function of electrochemical potential as reactant and product concentrations change, rather than mere temporal progression. \textbf{(2)} Battery interphases are complex solid-liquid mixtures with numerous components; also, their spectra combine overlapping, approximately linear contributions from many species, which greatly complicates interpretation. \textbf{(3)} Realistic spectral trajectories must satisfy fundamental physical laws such as mass conservation and peak shift.

To address these challenges, we propose \textbf{ABCC}, the first end-to-end operando IR prediction framework, featuring three core innovations designed to model these unique chemical and physical priors.
First, we introduce Chemical Flow, an explicit ordered representation of underlying reaction trajectories in spectral space. This is further modeled by MeanFlow through process averaging, avoiding autoregressive drift and robustly capturing the underlying evolution. 
Next, we propose a mixture-level embedding for solid-liquid systems alongside a Two-Stream Disentanglement mechanism, separately modeling solvent fluctuations and SEI chemical growth. 
Finally, we incorporate physics-informed constraints: Channel-wise Mean Alignment enforces mass conservation, while a peak constraint ensures realistic spectral shifts.
In short, our key contributions comprise:
\begin{itemize}[itemsep=2pt, topsep=0pt, parsep=0pt]
\item \textbf{A Large-Scale Operando Spectral Dataset}: We introduce the first curated benchmark containing 7,118 samples across 10 battery systems, accompanied by standardized evaluation protocols.
\item \textbf{A Novel ABCC Architecture}: Our model achieves noise-robust, trajectory-aware spectral transformation, accurately represents multi-component systems, and inherently respects known physical laws.
\item \textbf{Superior Empirical Performance}: The proposed method consistently surpasses state-of-the-art static, sequential, and video-based baselines, demonstrating strong chemical accuracy and generalization across diverse battery chemistries.
\end{itemize}

\section{Related Work}
\subsection{Related Datasets}
\cref{tab:dataset_comparison} summarizes several reported datasets for electrolyte IR spectroscopy along key dimensions:

\begin{table*}[htbp]
\centering
\caption{Comparison of electrolyte IR datasets: ours is a large-scale, open, ML-ready, sequential dataset with benchmark and evaluation.}
\vspace{-2mm}
\resizebox{2\columnwidth}{!}{%
\begin{tabular}{lcccccccc}
\toprule
\textbf{Dataset} & \textbf{Scale} & \textbf{Ratio Label} & \textbf{Openness} & \textbf{ML-Ready} & \textbf{Real-World} & \textbf{Sequential} & \textbf{Benchmark} & \textbf{Evaluation} \\
\midrule
\citet{Ellis2018A} & 81 & \cmark & \xmark & \cmark & \cmark & \xmark & \xmark & \xmark \\
\citet{Meyer2023Using} & 1,000 & \cmark & \xmark & \cmark & Simulation & \xmark & \xmark & \xmark \\
\citet{Buteau2019User-Friendly} & 40 & \cmark & \cmark & \cmark & \cmark & \xmark & \xmark & \xmark \\
\citet{Buteau2019User-Friendly} (robot) & 300 & \xmark & \cmark & \cmark & \cmark & \xmark & \xmark & \xmark \\
\citet{Kim2020Lithium} & 6 & \cmark & \xmark & \xmark & \cmark & \xmark & \xmark & \xmark \\
\midrule
\textbf{Ours} & \textbf{7,118} & \textbf{\cmark} & \textbf{\cmark} & \textbf{\cmark} & \textbf{\cmark} & \textbf{\cmark} & \textbf{\cmark} & \textbf{\cmark} \\
\bottomrule
\end{tabular}}
\label{tab:dataset_comparison}
\vspace{-0.5cm}
\end{table*}

\noindent\textbf{Dataset Size and Quality:} Existing datasets are generally small and often lack the time-resolved structure needed to model electrochemical dynamics. For example, \cite{Ellis2018A} reported 81 static spectra. \cite{Buteau2019User-Friendly} collected 40 manual samples and 300 robot-assisted ones, but the robot dataset does not provide precise composition labels. Our dataset contains 7,118 time-sequential spectra acquired from operando experiments, with precise multi-voltage annotations and systematic variation in salt concentration and mixture ratios. This design yields substantially finer experimental granularity and supports trajectory-aware machine learning for complex electrochemical evolution.

\noindent\textbf{Availability:} Data accessibility remains a major limitation in this area. Several prior collections \cite{Ellis2018A,Meyer2023Using,Kim2020Lithium} have not been publicly released, which limits reproducibility and slows comparative benchmarking. We release an open-access, ML-ready benchmark for operando electrolyte IR, accompanied by standardized preprocessing, predefined splits, and evaluation protocols. These components are intended to reduce implementation overhead and enable fair comparison in computational spectroscopy and battery informatics.

\noindent\textbf{Simulation v.s. Real-World:} Prior work also reflects a trade-off between simulation and experiment. Simulated spectra, such as the 1,000 interpolated spectra from \citet{Meyer2023Using}, provide controlled variability but may not reflect the full complexity of real measurements. Experimental datasets better capture realistic noise and artifacts but are often too small for robust learning. Our dataset provides large-scale experimental measurements collected under controlled laboratory conditions, including tightly managed potential, temperature, and electrolyte composition.

Although some operando electrolyte IR spectral data exist~\cite{Gervillié-Mouravieff2022Unlocking, Meyer2021Operando, Meyer2023Using, Leau2025Tracking}, they are often proprietary, typically limited in scale and variety, and lack systematic annotation, making them unsuitable for machine learning. We present the first open, ML-friendly dataset for operando IR spectroscopy, along with curated benchmark and evaluation protocols.

\subsection{Related Methods}
\noindent\textbf{Machine Learning in Electrolyte IR Spectroscopy:}
By using ML models, \citet{Ellis2018A,Buteau2019User-Friendly,Gervillié-Mouravieff2022Unlocking,Meyer2023Using} focus on analyzing static electrolyte component concentrations, \citet{yuan2025rapid} on analyzing the behavior of high-concentration Cu$^{2+}$, \citet{wei2020machine} on classification of amino acids and inorganic salts, and \citet{ishikawa2019machine} on predicting electrolyte properties. They apply chemometric tools, CNNs, and Beer's Law-based approaches to quantify lithium salt and solvent ratios rapidly and accurately. \citet{Meyer2023Using} also includes operando analysis, using PCA and CNNs to track lithium concentration. However, they do not address the challenging dynamic operando spectral prediction, and do not fully model complex spectral overlaps or enforce physical constraints as in the proposed method. Other works~\cite{Leau2025Tracking,meyer2021onoperando} explore operando IR spectroscopy to analyze electrolyte SEI dynamics, but often rely on multivariate curve resolution without integrated end-to-end predictive frameworks.

\noindent\textbf{Flow Matching:}
Flow Matching (FM)~\cite{lipman2022flow,albergo2022building,liu2022flow,albergo2023stochastic} resembles and builds upon diffusion models~\cite{ho2020denoising,song2020score}, and has been most widely developed and validated in image generation.
It learns a marginal instantaneous velocity field $\nu(z_t,t)=\mathbb{E}[\nu_t\mid z_t]$ from conditional targets $\nu_t=a_t'x+b_t'\varepsilon$, and generates samples by solving the probability-flow ODE $\frac{d z_t}{dt}=\nu(z_t,t)$.
As coarse discretizations poorly approximate curved marginal trajectories, one-step generation remains challenging. Few-step lines of work include consistency      models~\cite{yang2023weighted,song2025improved,geng2024consistency,lu2024simplifying} and distillations~\cite{salimans2022progressive,geng2023one,sauer2024adversarial,luo2023diff}, which enforce network-level self-consistency across times. MeanFlow~\cite{geng2025meanflowsonestepgenerative} reframes this by introducing an average-velocity field
\begin{equation}\label{eq:1}
\mathbf{u}(z_t,r,t)\triangleq \frac{1}{t-r}\int_r^t \nu(z_\tau,\tau)\,d\tau,
\end{equation}
and the MeanFlow identity $\mathbf{u}=\nu-(t-r)\frac{d}{dt}u$, yielding principled one-step objectives without distillation or curricula.
Traditional IR predictors miss structural heterogeneity, while FM suffers from integration errors and lacks multimodal conditioning and electrochemical supervision. Our ABCC conditions MeanFlow on system cues to learn average velocity fields. This enables one-step transport and effective modeling of realistic operando dynamics.

Additional related work, including \textbf{Physics-ML hybrids} and \textbf{Spectra generation}, is detailed in the \hyperref[appendix:relatedwork-hybrids]{\textit{Appendix}}.

\section{OpIRSpec-7K and OpIRBench}
\noindent\textbf{Data Collection and Processing.}
OpIRSpec-7K dataset designed for a reliable, system-level study for operando electrolyte IR spectroscopy. 
It covers 10 rigorously controlled electrolyte systems with distinct mixture recipes, both from ~\citet{Leau2025Tracking} and our experiments using a Bruker Vertex 70 Fourier-transform infrared spectrometer, providing a well-referenced coordinate for comprehensive learning across compositions.
Because trace electrolyte changes can strongly affect interfacial chemistry, we represent each electrolyte as an explicit mixture: a set of the salt and all solvents with their corresponding ratios, i.e., $electrolyte = \{molecule: proportion, …\}$. This formulation preserves compositional fidelity while remaining directly usable for machine learning.
The interphase evolution with high temporal resolution was captured by continuously recording spectra over multi-hour experiments with a chalcogenide specialty IR fiber, yielding 7,118 data points.
We further apply automated, community-standard preprocessing, including informative wavelength truncation, scale clipping, and normalization to mitigate global shifts. 
OpIRSpec-7K will be released as an open-source dataset to facilitate reproducible evaluation and method development.

\begin{figure}[t]
  \begin{center}
    \centerline{\includegraphics[width=0.85\columnwidth]{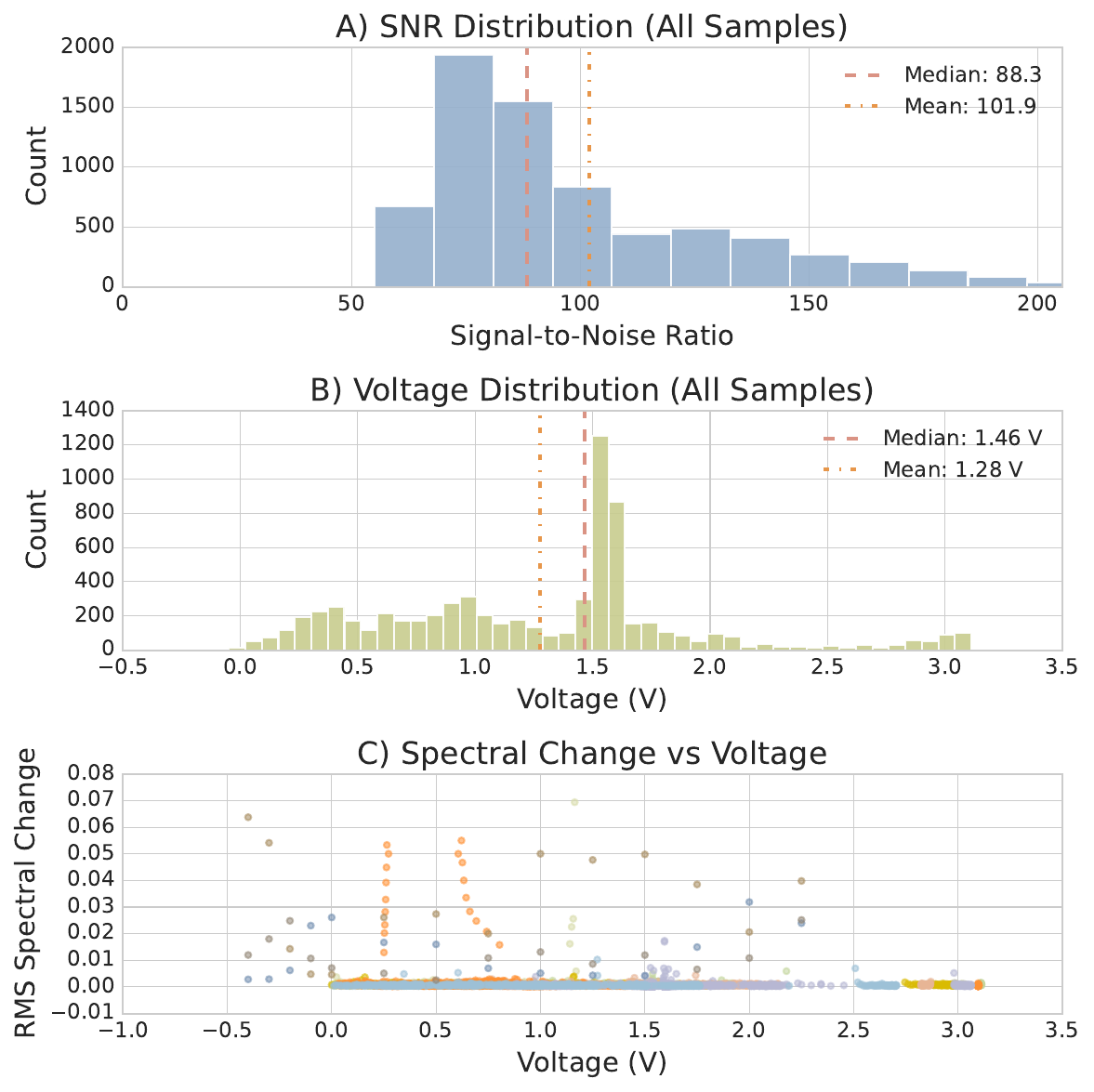}}
    \vspace{-2mm}
    \caption{OpIRSpec-7K shows A) high measurement fidelity with strong SNR, B) broad voltage coverage concentrated near operating potentials, and C) RMS spectral changes that combine subtle solvent-driven physical variation with sparse SEI chemical reaction-induced transitions, supporting high-quality operando data with chemically meaningful dynamics.}
    \label{fig:data_analysis_combined}
  \end{center}
\vspace{-0.5cm}
\end{figure}

\noindent\textbf{Data Analysis.}
\cref{fig:data_analysis_combined} shows the analysis of OpIRSpec-7K data quality through complementary distribution and dynamics.
\textbf{SNR histogram} indicates consistently strong measurement fidelity across all spectra, with a median SNR of $88.3$ and a mean of $101.9$. Most samples concentrate in the high SNR regime, and the right-tailed distribution suggests that many experiments achieve exceptionally clean signals. This supports robust learning, where models benefit from informative spectral variation with limited stochastic noise.
The \textbf{voltage histogram} summarizes electrochemical coverage during operando acquisition. Samples span a broad potential window, with dense sampling around the primary operating region between $0.2$ to $2.0$. The combination of wide coverage and concentrated sampling improves statistical power for voltage-conditioned analyses.
Finally, the \textbf{RMS spectral change vs. voltage} quantifies temporal dynamics. Most points cluster near zero, indicating stable acquisition, minimal artifacts, and effective preprocessing. It also highlights operando dynamics: \textbf{Most spectra show small solvent-driven physical variations, punctuated by sparse jump-like events of SEI chemical reactions}. This mixed pattern indicates stable acquisition while preserving chemically meaningful transitions. Thus, OpIRSpec-7K provides high-quality, reproducible trajectories suitable for system-level modeling of SEI growth.

\begin{algorithm}[t]
\caption{Spectrum-Waveform Auto-Conversion}\label{alg:spec-img}
\begin{algorithmic}
\Function{SpecToImage}{$\mathbf{s},\ H,\ W,\ \sigma$}
\State \textbf{Inputs:} spectrum $\mathbf{s}\in\mathbb{R}^{N_f}$, image height $H\in\mathbb{N}$, image width $W\in\mathbb{N}\in\mathbb{N}$, Gaussian width $\sigma\in\mathbb{R}_{+}$
\State \textbf{Constants:} $s_{\max}\in\mathbb{R}_{+}$ (normalizer), $m\in[0,0.5)$ (vertical margin), $g\in\mathbb{R}_{+}$ (contrast)
\State \textbf{Outputs:} grayscale image $\mathbf{I}\in\mathbb{R}^{H\times W}$
\State $\mathbf{s}\gets\text{clip}(\mathbf{s}/s_{\max},-1,1)\in\mathbb{R}^{N_f}$
\State $\mathbf{s}_r\gets\text{resample}(\mathbf{s},W)\in\mathbb{R}^{W}$
\State $\alpha\gets\frac{H(1-2m)}{2}\in\mathbb{R}_{+}$;\quad $\mathbf{y}\gets\frac{H-1}{2}-\alpha\mathbf{s}_r\in\mathbb{R}^{W}$
\State $I_{i,j}\gets\exp\!\left(-\frac{(i-y_j)^2}{2\sigma^2}\right),\ \forall i\in\{0,\dots,H-1\},\ j\in\{0,\dots,W-1\}$
\State \Return $255-g\mathbf{I}$
\EndFunction
\\
\Function{ImageToSpec}{$\mathbf{I},\ N_f,\ \alpha,\ s_{\max}\in\mathbb{R}_{+}$}
\State \textbf{Inputs:} grayscale image $\mathbf{I}\in\mathbb{R}^{H\times W}$, target length $N_f\in\mathbb{N}$, scaling $\alpha\in\mathbb{R}_{+}$, normalizer $s_{\max}\in\mathbb{R}_{+}$
\State \textbf{Constants:} $g\in\mathbb{R}_{+}$ (contrast)
\State \textbf{Outputs:} reconstructed spectrum $\hat{\mathbf{s}}\in\mathbb{R}^{N_f}$
\State $\mathbf{E}\gets(255-\mathbf{I})/g\in\mathbb{R}^{H\times W}$
\State $\hat{y}_j\gets\frac{\sum_{i=0}^{H-1} iE_{i,j}}{\sum_{i=0}^{H-1} E_{i,j}}\in\mathbb{R},\ \forall j\in\{0,\dots,W-1\}$
\State $\hat{\mathbf{s}}_r\gets\left(\frac{H-1}{2}-\hat{\mathbf{y}}\right)/\alpha\in\mathbb{R}^{W}$
\State \Return $\text{resample}(\hat{\mathbf{s}}_r,N_f)\cdot s_{\max}$
\EndFunction
\end{algorithmic}
\end{algorithm}

\noindent\textbf{Evaluation Metrics.}
We use MAE, MSE, RMSE to measure pointwise reconstruction fidelity, Spectral Angle Mapper (SAM) \& Cosine Similarity to assess spectral shape consistency and scale-invariance, $R^2$ Coefficient to quantify explained variance across wavenumbers, and DILATE~\cite{9721108} to evaluate temporal coherence and the timing of abrupt dynamic transitions. 

\noindent\textbf{Spectrum-Waveform Bidirectional Auto-Conversion.}
To leverage mature image-based generative models while keeping evaluation in the spectral domain, we design a fast, reversible conversion between a 1D spectrum $\mathbf{s}\in\mathbb{R}^{N_f}$ and a 2D grayscale waveform image $\mathbf{I}\in\mathbb{R}^{H\times W}$. 
Algorithm~\ref{alg:spec-img} shows the forward mapping (Spec$\to$Img) and the inverse mapping (Img$\to$Spec).
This bidirectional design is lightweight, numerically stable, and enables plug-and-play use for these models without losing physical interpretability.

\noindent\textbf{OpIRBench Benchmark.}
Due to the limited length, details of \textbf{Problem Definition}, \textbf{Metrics}, and \textbf{Dataset Split (\textit{i.e.}, Random and System Split)} are provided in the \hyperref[appendix:OpIRBench]{\textit{Appendix}}.


\section{Method}


\begin{figure*}[h]
    \centering
    \includegraphics[width=0.8\linewidth]{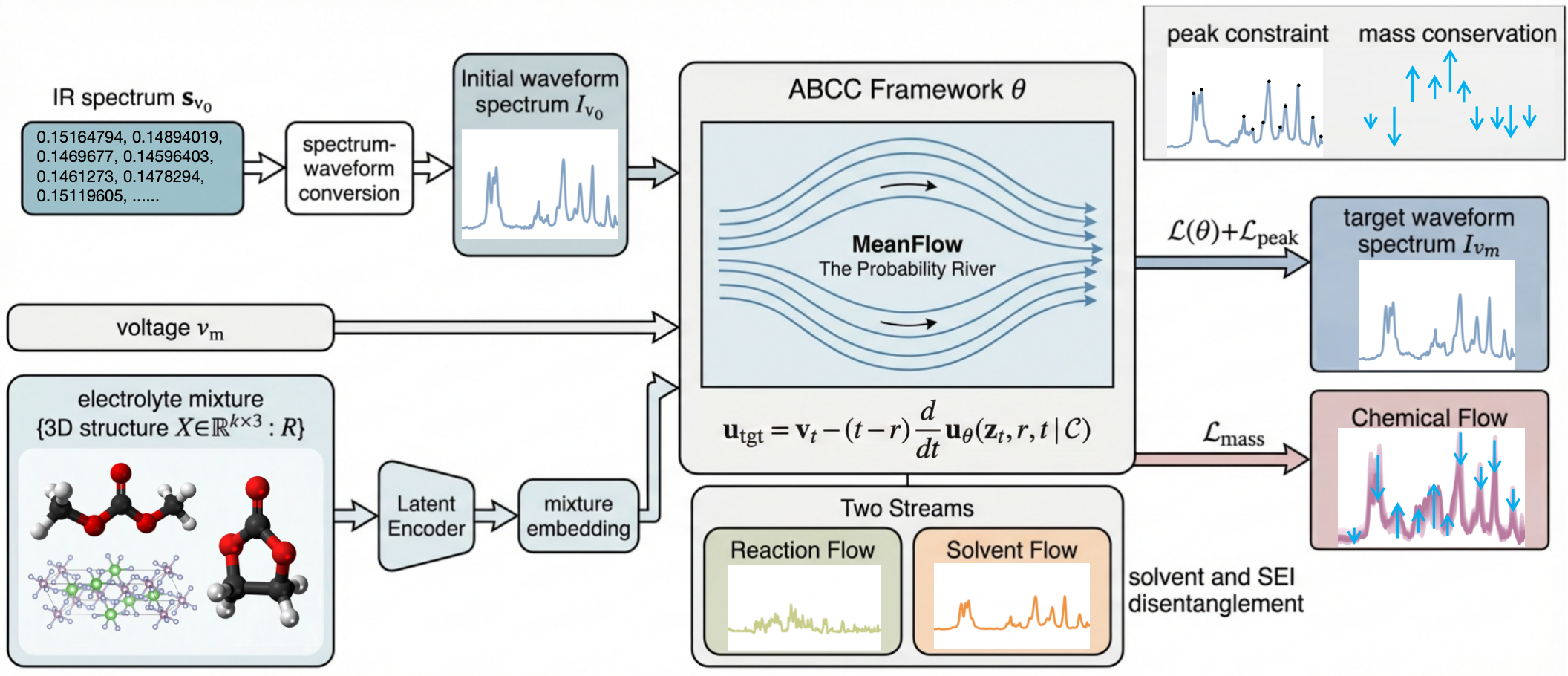}
    \vspace{-2mm}
    \caption{Training pipeline of the proposed ABCC framework. The input IR spectrum $\mathbf{s}_{v_0}$ is converted to an initial waveform spectrum $I_{v_0}$ and fed into ABCC together with the voltage condition $v_m$. The electrolyte mixture, represented by 3D molecular structures $X\in\mathbb{R}^{k\times 3}$ and their ratios $R$, is encoded by a latent encoder to produce a mixture embedding used as an additional condition $\mathcal{C}$. ABCC models the conditional probability river of MeanFlow and is optimized by matching the target waveform spectrum $I_{v_m}$ with the objective $\mathcal{L}(\theta)$. Two-stream design separates reaction flow and solvent flow to promote solvent and SEI disentanglement.  A peak constraint $\mathcal{L}_{\mathrm{peak}}$ and a mass conservation loss $\mathcal{L}_{\mathrm{mass}}$ incorporate physics-informed constraints. At inference, ABCC forecasts full operando IR spectral dynamics from a single static spectrum, conditioned on voltage profiles and electrolyte composition, generalizing to unseen systems.}

    \label{fig:method}
\vspace{-0.3cm}
\end{figure*}

\noindent\textbf{MeanFlow for Operando IR Prediction.}
To capture chemically driven peak shifts and nonstationary band evolution in operando IR spectroscopy, we cast spectrum prediction as a sequence-conditioned generative process. Classical spectroscopy forward modeling often uses deterministic regression that maps experimental context to a single continuous spectrum~\cite{ABDULAL2024141603,10.24963/ijcai.2025/1160,mcgill2021predicting,Ren2021A}, which can underrepresent structural heterogeneity and stochastic fluctuations during electrolyte decomposition. We address this by repurposing the predictor into a conditional MeanFlow generator. Concretely, the 2D waveform image $I_{v_m}$ obtained via our auto-conversion is encoded into a VAE latent $\mathbf{z}_{v_m}=E(I_{v_m})$, yielding a compact manifold for generative learning. This provides a general mechanism for turning conditional regression into conditional generation.

We define a conditional probability path $p_t(\mathbf{z}\mid \mathbf{z}_{v_m})$ that bridges a standard Gaussian prior $q(\boldsymbol{\epsilon})=\mathcal{N}(\mathbf{0},\mathbf{I})$ and the target latent distribution concentrated at $\mathbf{z}_{v_m}$, where $z$ is the interpolated lantent state. We use a linear interpolation map to specify the prediction process,
\begin{equation}
\phi_t(\mathbf{z}_{v_m}) = (1-t)\mathbf{z}_{v_m} + t\,\boldsymbol{\epsilon}, 
\qquad \boldsymbol{\epsilon}\sim\mathcal{N}(\mathbf{0},\mathbf{I}),\quad t\in[0,1],
\end{equation}
which induces the conditional instantaneous velocity field
\begin{equation}
\mathbf{\nu}_t(\mathbf{z}_t\mid \mathbf{z}_{v_m}) = \boldsymbol{\epsilon}-\mathbf{z}_{v_m},
\end{equation}
where $t$ represents the internal flow time. 
Rather than training a network to approximate the instantaneous marginal velocity field as in FM, we follow MeanFlow and learn an average velocity field that summarizes the net displacement over a time interval. For two time points $r<t$, the MeanFlow average velocity is defined in~\cref{eq:1}. 

Our conditional MeanFlow generator is parameterized as an average-velocity field
$\mathbf{u}_\theta(\mathbf{z}_t,r,t\mid \mathcal{C})$, where $\mathcal{C}$ includes the initial template
$\mathbf{z}_{v_0}=E(I_{v_0})$, the driving voltage $v_m$, and the electrolyte descriptor
$\{X\!:\!R,\ldots\}$. MeanFlow training is guided by the intrinsic identity that relates average velocity to the instantaneous velocity:
\begin{equation}
\mathbf{u}(\mathbf{z}_t,r,t)=\mathbf{\nu}_t(\mathbf{z}_t,t)-(t-r)\frac{d}{dt}\mathbf{u}(\mathbf{z}_t,r,t).
\end{equation}
We replace the intractable marginal instantaneous velocity with the sample-conditional velocity on the
linear interpolation path, $\mathbf{\nu}_t=\boldsymbol{\epsilon}-\mathbf{z}_{v_m}$, and compute the total
derivative via a Jacobian-vector product with tangent $(\mathbf{\nu}_t,0,1)$:
\begin{equation}
\frac{d}{dt}\mathbf{u}_\theta(\mathbf{z}_t,r,t\mid \mathcal{C})
=
\mathbf{\nu}_t^\top \partial_{\mathbf{z}}\mathbf{u}_\theta(\mathbf{z}_t,r,t\mid \mathcal{C})
+\partial_t\mathbf{u}_\theta(\mathbf{z}_t,r,t\mid \mathcal{C}).
\end{equation}
This yields the MeanFlow target and loss:
\begin{equation}
\mathbf{u}_{\mathrm{tgt}}=\mathbf{\nu}_t-(t-r)\frac{d}{dt}\mathbf{u}_\theta(\mathbf{z}_t,r,t\mid \mathcal{C}), 
\end{equation}
\begin{equation}
\mathcal{L}(\theta)=\mathbb{E}\big\|\mathbf{u}_\theta-\mathrm{sg}(\mathbf{u}_{\mathrm{tgt}})\big\|_2^2.
\end{equation}
At inference, we perform direct transport using the learned average velocity,
\begin{equation}
\mathbf{z}_r=\mathbf{z}_t-(t-r)\mathbf{u}_\theta(\mathbf{z}_t,r,t\mid \mathcal{C}),
\end{equation}
and in one step
\begin{equation}
\hat{\mathbf{z}}_{v_m}=\boldsymbol{\epsilon}-\mathbf{u}_\theta(\boldsymbol{\epsilon},0,1\mid \mathcal{C}),
\ \boldsymbol{\epsilon}\sim\mathcal{N}(\mathbf{0},\mathbf{I}).
\end{equation}
This formulation avoids error accumulation in auto-regressive forecasting and produces a conditional distribution over plausible trajectories, improving robustness and fidelity to peak shifts and intensity variations across wavenumbers. It is also more efficient than FM by enabling single-step transport from noise to target, mitigating integration errors when latent trajectories are curved.

\begin{algorithm}[t]
\caption{Chemical Flow image construction}\label{alg:chemflow}
\begin{algorithmic}
\Function{SpecToCFlow}{$\{\mathbf{s}_{v_t}\}_{t=0}^T, H, W, m, \alpha$}
\State \textbf{Inputs:} spectra sequence $\{\mathbf{s}_{v_t}\}_{t=0}^T$, image size $(H,W)$, margin $m$, clip scale $\alpha$
\State \textbf{Outputs:} $\mathbf{I_{CF}}\in\mathbb{R}^{H\times W}$
\State initialize Chemical Flow image $\mathbf{I_{CF}}\leftarrow 255\cdot \mathbf{1}_{H\times W}$, $\mathbf{s}_{\mathrm{ref}}\leftarrow \mathbf{s}_{v_0}$
\For{$t=0$ to $T$}
\State $\Delta \mathbf{s}\leftarrow \mathbf{s}_{v_t}-\mathbf{s}_{\mathrm{ref}}$
\State $\mathbf{x}\leftarrow \mathrm{clip}(\Delta \mathbf{s},-\alpha,\alpha)/\alpha$ \Comment{normalize to $[-1,1]$}
\State $\mathbf{x}\leftarrow \mathrm{Resample}(\mathbf{x},W)$
\State $\mathbf{y}\leftarrow \mathrm{Centerline}(H)-\mathbf{x}\cdot (1-2m)\cdot (H/2)$
\State $g_t \leftarrow 200\cdot\left(1-\frac{t}{\max(1,T)}\right)$ \Comment{light gray $\rightarrow$ black}
\State draw polyline $(0{:}W{-}1,\mathbf{y})$ with grayscale $g_t$
\EndFor
\State \Return $\mathbf{I_{CF}}$
\EndFunction
\end{algorithmic}
\end{algorithm}

\noindent\textbf{Chemical Flow and MeanFlow.}
To align with MeanFlow’s emphasis on global dynamics, we introduce Chemical Flow, a trajectory-level representation of operando spectral evolution. Given a voltage-indexed sequence of spectra $\{\mathbf{s}_{v_0},\mathbf{s}_{v_1},\ldots,\mathbf{s}_{v_M}\}$, we define the chemical variation field relative to the reference state $v_0$ as
\begin{equation}
\Delta \mathbf{s}(v_m)=\mathbf{s}{v_m}-\mathbf{s}{v_0},\qquad t=0,\ldots,M.
\end{equation}
Our innovations are three-fold: First, by geometrically stacking $\{\Delta \mathbf{s}(v_m)\}_{t=0}^M$, we yield a family of \textbf{trajectories} starting from the origin in variation space, which makes the representation consistent with a flow viewpoint where voltage serves as the trajectory parameter. Second, we \textbf{explicitly encode order} through grayscale, forming a voltage-parameterized trajectory in a 2D canvas rather than attempting to recover instantaneous dynamics.
Third, to mirrors MeanFlow’s denoising-by-averaging perspective and further \textbf{stabilize} learning, we apply clipping and normalization,
which projects high-dimensional, noisy spectral fluctuations onto a bounded, low-variance visible submanifold. 
The Chemical Flow construction is described in Algorithm~\ref{alg:chemflow}. In all, our Chemical Flow provides a trajectory-level projection of spectral evolution relative to a reference state, effectively visualizing the underlying spectral flow trajectory.

\paragraph{Two-Stream Disentanglement and Dual-Flow Model.}
Operando IR spectra measured near battery electrodes conflate two sources of variation with distinct dynamics. The \textbf{solvent physical movement} evolves through reversible solvation and concentration gradients induced by polarization, in contrast with \textbf{SEI chemical formation} that follows largely irreversible trajectories and introduces new carbonate and semi-carbonate absorptions. Thus, we propose a two-stream architecture that disentangles solvent fluctuations from SEI growth dynamics, improving interpretability as well as transfer and robustness under shifts in salt concentration, solvent ratio, and electrode material.
In preprocessing, we decompose the spectrum $D(v)$ with:
\begin{equation}
D(v)=C_{\text{elec}}(v)S_{\text{elec}}^{\mathsf T}+C_{\text{SEI}}(v)S_{\text{SEI}}^{\mathsf T}+E,
\end{equation}
where $C_{\text{elec}}(v)$ and $C_{\text{SEI}}(v)$ are concentration profile matrix and ideally measurement noise $E \to 0$. We note that, to prevent data leakage, electrolyte spectra $S_{\text{elec}}$ and SEI spectra $S_{\text{SEI}}$ are learned from another reference dataset of static calibration~\cite{Leau2025Tracking}, fixed during our training and inference.
We then adopt a dual-flow Transformer with two dedicated branches $\mathcal{F}_{\text{elec}}$ and $\mathcal{F}_{\text{SEI}}$ that model solvent and SEI, respectively. Representations are fused via learnable weights
for physically interpretable reconstruction.

\noindent\textbf{3D Molecular Mixture Representation.}  
Peak positions and intensities shift with Li$^+$ coordination, ion pairing, and competitive solvation, and these effects vary across solvent ratios, salt molarity, and additives, yielding a combinatorial mixture space. However, general end-to-end regressors that map a mixture of \textbf{multiple molecular} to a \textbf{single} operando spectrum remain rare. To provide a transferable condition, we encode the 3D geometry of each component with Uni-Mol~\cite{zhou2023unimol}, then aggregate them into a mixture embedding
$\mathbf{c} = [\mathbf{e}_{\text{salt}}, \sum_{i} p_i \mathbf{e}_i],$
where $\mathbf{e}_{\text{salt}},\mathbf{e}_i\in\mathbb{R}^{512}$, $\mathbf{e}_i$ are solvents and $p_i$ are their normalized volume fractions.

\begin{table*}[t]
\centering
\small
\setlength{\tabcolsep}{4pt}
\caption{Random Split quantitative evaluation on OpIRBench. Numbers are reported as mean$\pm$std with scaling. Arrows indicate the preferred direction. Metrics are computed on 1D spectra. Our ABCC dominates point-wise, shape, and dynamics metrics.}
\vspace{-2mm}
\resizebox{2\columnwidth}{!}{
\begin{tabular}{lccccccc}
\toprule
Method &
MAE ($\times 10^{-2}$)$\downarrow$ &
MSE ($\times 10^{-3}$)$\downarrow$ &
RMSE ($\times 10^{-2}$)$\downarrow$ &
SAM($^\circ$)$\downarrow$ &
CosSim$\uparrow$ &
$R^2$$\uparrow$ &
DILATE$\downarrow$ \\
\midrule
\multicolumn{8}{l}{\textit{IR spectrum generation (static):}}\\
\rowcolor{IRCol}
NNMol-IR     & $11.21\err{1.04}$ & $20.3\err{2.9}$ & $14.18\err{1.32}$ & $89.26\err{6.66}$ & $0.013\err{0.115}$ & $-3.89\err{2.74}$ & $4.11$ \\
\rowcolor{IRCol}
MACE4IR      & $8.98\err{1.15}$  & $13.7\err{3.3}$ & $11.62\err{1.42}$ & $89.38\err{9.12}$ & $0.011\err{0.167}$ & $-2.28\err{1.90}$ & $3.37$ \\
\midrule
\multicolumn{8}{l}{\textit{Video generation:}}\\
\rowcolor{VideoCol}
CogVideoX     & $7.77\err{1.16}$  & $10.9\err{2.8}$ & $10.33\err{1.36}$ & $90.74\err{9.40}$ & $-0.013\err{0.162}$ & $-1.52\err{0.68}$ & $3.00$ \\
\rowcolor{VideoCol}
Pyramid-Flow  & $5.67\err{0.85}$  & $6.8\err{1.5}$  & $7.65\err{0.93}$  & $51.29\err{5.64}$ & $0.573\err{0.073}$ & $-0.70\err{2.04}$ & $2.22$ \\
\midrule
\multicolumn{8}{l}{\textit{Time-series forecasting:}}\\
\rowcolor{TSCol}
STDN          & $7.47\err{1.19}$  & $10.3\err{2.7}$ & $10.04\err{1.36}$ & $91.12\err{9.25}$ & $-0.019\err{0.159}$ & $-1.36\err{0.47}$ & $2.91$ \\
\midrule
\multicolumn{8}{l}{\textit{Reference:}}\\
\rowcolor{RefCol}
Random        & $13.06\err{1.35}$ & $25.3\err{3.8}$ & $15.89\err{1.41}$ & $92.97\err{9.73}$ & $-0.052\err{0.162}$ & $-5.19\err{4.31}$ & $4.61$ \\
\midrule
\rowcolor{OursCol}
Ours & $\mathbf{1.11\err{0.38}}$ & $\mathbf{0.24\err{0.17}}$ & $\mathbf{1.49\err{0.44}}$ & $\mathbf{8.77\err{4.56}}$ & $\mathbf{0.986\err{0.043}}$ & $\mathbf{0.93\err{0.19}}$ & $\mathbf{0.43}$ \\
\bottomrule
\end{tabular}}
\label{tab:random_split}
\vspace{-0.3cm}
\end{table*}

\noindent\textbf{Channel-wise Mean Alignment and Peak Constraint.} 
To encourage physically plausible difference spectra, we add two auxiliary losses. First, the mass conservation enforced by \textbf{channel-wise mean alignment}: generating SEI products consumes a corresponding amount of electrolyte, so the predicted difference spectrum should contain both positive and negative changes. We therefore match the global mean
\begin{equation}
\mathcal{L}{\text{mass}}=\left|\bar{u}{\mathbf{I^{pred}_{CF}}}-\bar{u}{\mathbf{I_{CF}}}\right|,\qquad
\bar{u} = \mathbb{E}_{h,w}[u_{h,w}],
\end{equation}
where $\mathbf{I^{pred}_{CF}}$ is the predicted Chemical Flow, $\bar{u}$ computes the average pixel value for each channel. This constraint balances product growth and solvent depletion across the reconstructed image.

Second, operando spectra often exhibit prominent peaks whose amplitudes evolve with reaction progress, while their positions may remain stable for some modes but can also drift due to physically meaningful effects. Thus, instead of enforcing strict positional invariance, we regularize the model to reproduce the relative peak displacement, by a \textbf{peak constraint}:
\begin{equation}
\mathcal{L}{\text{peak}}=\sum{i=1}^{N}\left|\big(\Delta p_i^{\text{pred}}(t),a_i^{\text{pred}}\big)-\big(\Delta p_i(t),a_i\big)\right|_2^2,
\end{equation}
where $p_i\in[0,1]$ is the normalized peak position extracted from the target spectrum using scipy~\cite{virtanen2020scipy} prominence-based detection~\cite{Helman2005FinestPeaks} and $a_i$ is the peak area via trapezoidal~\cite{press2007numerical}, $\Delta p_i(t)=p_i(t)-p_i(0),\qquad 
\Delta p_i^{\text{pred}}(t)=p_i^{\text{pred}}(t)-p_i^{\text{pred}}(0)$. This regularizer favors vertical evolution of peaks and discourages spurious global horizontal drift.

More ABCC \textbf{Implementation Details} are in the \hyperref[appendix:ABCC]{\textit{Appendix}}.

\section{Experiments}
\noindent\textbf{Baselines.} 
We benchmark against IR Spectrum Generation (NNMol-IR \cite{ABDULAL2024141603}, MACE4IR \cite{bhatia2025mace4irfoundationmodelmolecular}) to evaluate static molecular responses, Video Generation (CogVideoX \cite{yangcogvideox}, Pyramid-Flow \cite{jinpyramidal}) to model long-range trajectory correlations, Time-Series Forecasting (STDN \cite{cao2025spatiotemporal}) to capture structured spatiotemporal variations, and Random Spectra to establish a performance lower bound. Details regarding these baselines are provided in the \hyperref[appendix:baselines]{\textit{Appendix}}.

\noindent\textbf{Quantitative Results on Random Split.}
\cref{tab:random_split} shows quantitative results for all baselines on Random Split of OpIRBench. Our proposed ABCC dominates all metrics by a large margin, with an order of magnitude lower pointwise errors, better shape agreement and dynamics. The random baseline is consistently the weakest, confirming that the benchmark is not solvable by marginal spectrum statistics. Static IR spectrum generation baselines, NNMol-IR and MACE4IR, perform poorly on DILATE, indicating limited ability to follow voltage-driven dynamics in the operando IR spectrum. Temporal generative models improve DILATE substantially over static IR models, indicating their ability to capture dynamics. Notably, Pyramid-Flow provides the strongest baseline across both framewise and temporal metrics, consistent with flow matching offering more stable sequence modeling and better dynamics and long-range coherence than the diffusion-based video baseline.

\begin{figure*}[t]
\centering
\small
\setlength{\tabcolsep}{4pt}
\captionof{table}{System Split quantitative evaluation. Our ABCC still dominates metrics, and shows strong generalization to unseen systems.}
\vspace{-2mm}
\resizebox{2\columnwidth}{!}{
\begin{tabular}{lccccccc}
\toprule
Method &
MAE ($\times 10^{-2}$)$\downarrow$ &
MSE ($\times 10^{-3}$)$\downarrow$ &
RMSE ($\times 10^{-2}$)$\downarrow$ &
SAM($^\circ$)$\downarrow$ &
CosSim$\uparrow$ &
$R^2$$\uparrow$ &
DILATE$\downarrow$ \\
\midrule
\multicolumn{8}{l}{\textit{IR spectrum generation (static):}}\\
\rowcolor{IRCol}
NNMol-IR     & $11.94\err{0.86}$ & $23.1\err{3.1}$ & $15.15\err{1.04}$ & $89.18\err{6.57}$ & $0.014\err{0.114}$ & $-3.21\err{0.67}$ & $4.40$ \\
\rowcolor{IRCol}
MACE4IR      & $10.02\err{1.02}$  & $16.9\err{3.0}$ & $12.93\err{1.14}$ & $89.94\err{9.82}$ & $0.010\err{0.169}$ & $-2.08\err{0.56}$ & $3.75$ \\
\midrule
\multicolumn{8}{l}{\textit{Video generation:}}\\
\rowcolor{VideoCol}
CogVideoX     & $8.89\err{0.76}$ & $13.8\err{1.8}$ & $11.71\err{0.77}$ & $90.93\err{9.37}$ & $-0.016\err{0.161}$ & $-1.51\err{0.32}$ & $3.40$ \\
\rowcolor{VideoCol}
Pyramid-Flow  & $6.91\err{0.48}$ & $8.9\err{1.8}$ & $9.10\err{0.49}$ & $52.76\err{4.87}$ & $0.56\err{0.072}$ & $-0.62\err{0.15}$ & $2.64$ \\
\midrule
\multicolumn{8}{l}{\textit{Time-series forecasting:}}\\
\rowcolor{TSCol}
STDN          & $8.64\err{0.69}$ & $13.1\err{1.5}$ & $11.44\err{0.67}$ & $91.04\err{9.31}$ & $-0.018\err{0.16}$ & $-1.40\err{0.27}$ & $3.32$ \\
\midrule
\multicolumn{8}{l}{\textit{Reference:}}\\
\rowcolor{RefCol}
Random        & $14.01\err{1.63}$ & $28.6\err{3.7}$ & $16.87\err{1.23}$ & $93.12\err{12.98}$ & $-0.058\err{0.19}$ & $-4.23\err{1.90}$ & $4.90$ \\
\midrule
\rowcolor{OursCol}
Ours          & $\mathbf{2.43\err{0.82}}$ & $\mathbf{1.12\err{1.06}}$ & $\mathbf{3.21\err{0.94}}$ & $\mathbf{14.73\err{6.61}}$ & $\mathbf{0.96\err{0.08}}$ & $\mathbf{0.80\err{0.17}}$ & $\mathbf{0.93}$ \\
\bottomrule
\end{tabular}}
\label{tab:system_split}
\setlength{\tabcolsep}{3pt}
\begin{tabular}{ccccccc}
\includegraphics[width=0.13\textwidth]{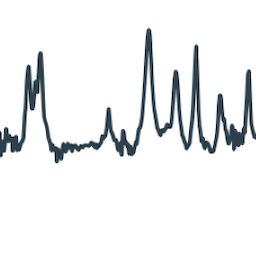} &
\includegraphics[width=0.13\textwidth]{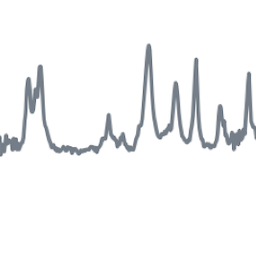} &
\includegraphics[width=0.13\textwidth]{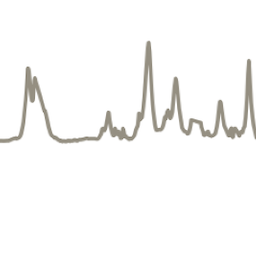} &
\includegraphics[width=0.13\textwidth]{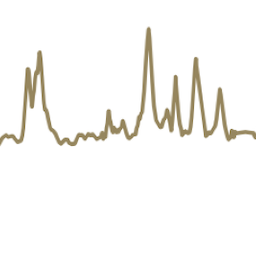} &
\includegraphics[width=0.13\textwidth]{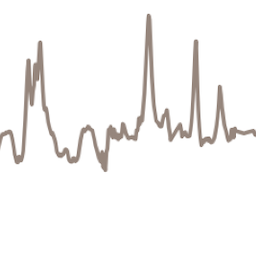} &
\includegraphics[width=0.13\textwidth]{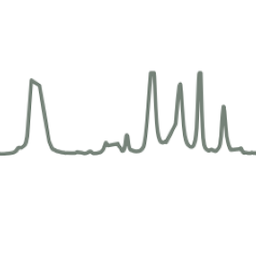} &
\includegraphics[width=0.13\textwidth]{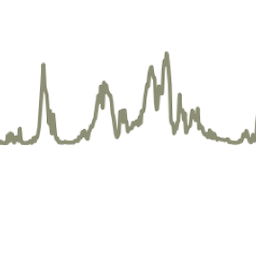} \\[-25pt]
\footnotesize Ground Truth (GT) &
\footnotesize Ours &
\footnotesize Pyramid-Flow &
\footnotesize STDN &
\footnotesize CogVideoX &
\footnotesize MACE4IR &
\footnotesize NNMol-IR \\
\end{tabular}
\captionof{figure}{Qualitative comparison on the System Split test set. From left to right: ground truth, ours, and competing baselines. Each panel visualizes the 1D IR spectrum rendered as a 2D colored plot for readability. Quantitative metrics in \cref{tab:system_split} are computed on 1D spectra, not on the rendered images. Our ABCC most closely matches peak locations and relative intensity, showing the best quality.}
\label{fig:systemsplit_qual}
\end{figure*}

\noindent\textbf{Quantitative and Qualitative Results on System Split.}
\cref{tab:system_split} shows the quantitative results on System Split.
Overall, the ranking largely mirrors Random Split. Static IR baselines show limited temporal alignment quality, with high DILATE and near orthogonal spectral angles. Video generation and time series forecasting baselines reduce DILATE relative to static IR models, and Pyramid-Flow is consistently stronger than CogVideoX, indicating an advantage of flow matching methods in modeling dynamics.
All methods degrade under System Split, confirming the generalization gap that extrapolating to unseen electrolyte systems is harder than predicting within the seen system distribution. The degradation is especially obvious in sequence metrics, which may be due to reaction timing and spectral evolution pattern changes across different systems. ABCC remains the top method across every metric, indicating its best generalization.  This suggests ABCC captures voltage-conditioned dynamics in a way that transfers across electrolyte compositions, rather than fitting system-specific spectral templates.
\cref{fig:systemsplit_qual} shows ABCC most closely matches the GT peak locations and relative intensities, while maintaining a smooth but sharp spectral profile. This aligns with \cref{tab:system_split}, where ABCC achieves the best quality and generalization.

\begin{figure}[t]
\centering
\small
\setlength{\tabcolsep}{2pt}
\captionof{table}{Ablation under system split. MAE is in $10^{-2}$, MSE in $10^{-3}$, $R^2$ in $10^{-1}$. Chemical Flow, two-stream disentanglement, 3D mixture, and physics constraints drive generalization.}
\vspace{-2mm}
\resizebox{\columnwidth}{!}{
\begin{tabular}{lccccc}
\toprule
Variant & MAE $\downarrow$ & MSE $\downarrow$ & SAM $\downarrow$ & $R^2$ $\uparrow$ & DILATE$\downarrow$ \\
\midrule
ABCC (full) & 2.4 & 1.1 & 14.7 & 8.0 & 0.93 \\
w/o Chemical Flow & 4.8 & 4.7 & 27.9 & 1.4 & 1.87 \\
\quad  + w/o Two-Stream & 5.4 & 6.4 & 36.9 & -1.6 & 1.98 \\
\quad + SMILES over 3D & 5.7 & 6.9 & 40.6 & -2.5 & 2.11 \\
\quad + w/o Peak Const. & 5.6 & 6.6 & 44.9 & -1.9 & 2.12 \\
\quad + w/o Mean Align. & 6.2 & 7.7 & 43.5 & -4.1 & 2.24 \\
\bottomrule
\end{tabular}}
\label{tab:ablation_system}
    \centering
    \includegraphics[width=0.9\linewidth]{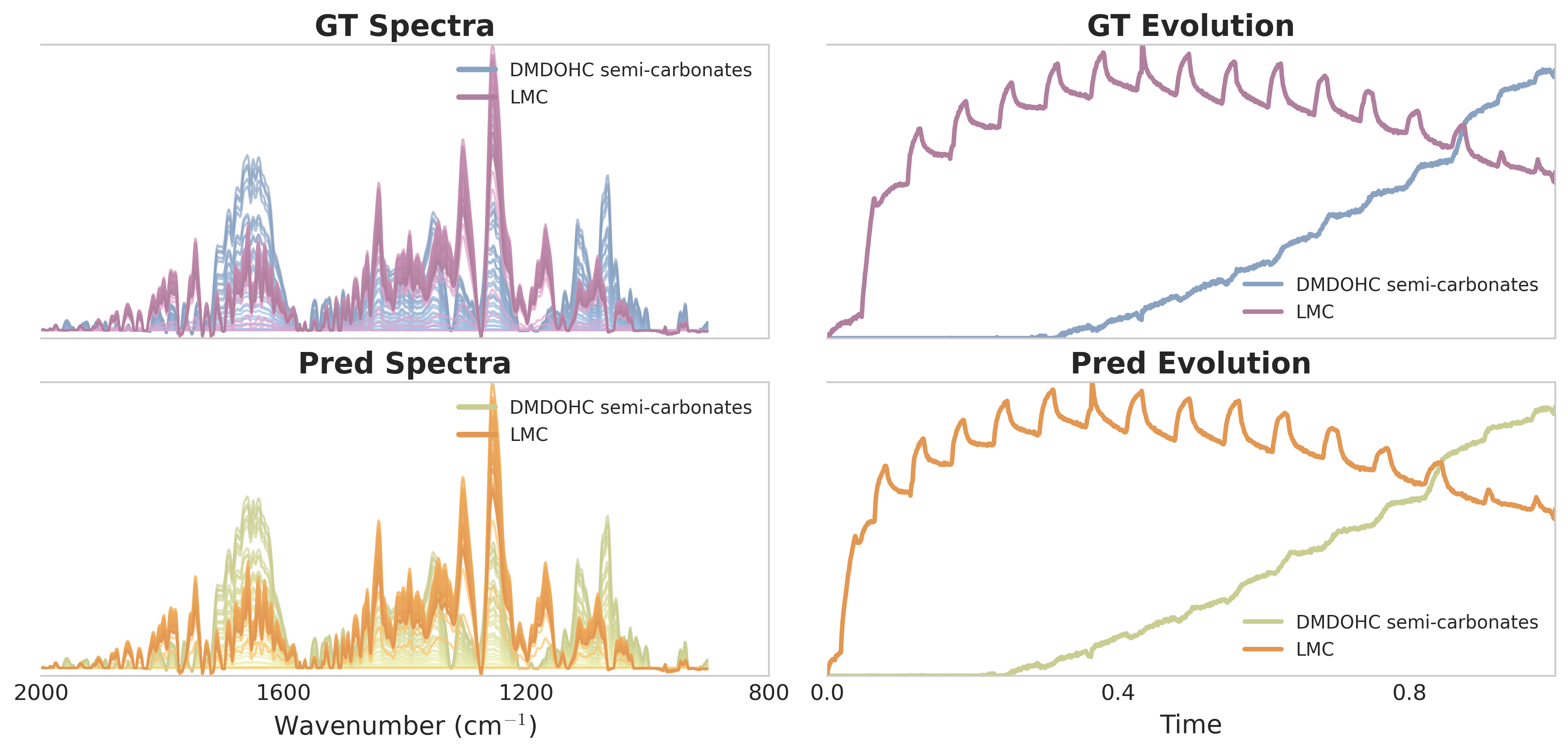}
    \captionof{figure}{We run the same SEI component evolution decomposition algorithm on both ground-truth operando spectra and on the spectrum sequence predicted by our model, and see similar results, aligning with the literature-reported trend in \cite{Leau2025Tracking}.}
    \label{fig:application}
\vspace{-0.3cm}
\end{figure}

\noindent\textbf{Ablation.}
\cref{tab:ablation_system} shows different components of our method under the System Split of OpIRBench. Removing Chemical Flow nearly doubles DILATE from $0.93$ to $1.87$ and degrades all spectrum level metrics, indicating that this trajectory level representation of operando evolution is critical for matching global dynamics and temporal alignment across systems.
Two Stream Disentanglement provides an additional, consistent gain. After removing it, the spectral angle increases from $27.9$ to $36.9$, and $R^2$ drops below zero, suggesting the model starts to mix solvent physical movements with system-specific SEI chemical formation, harming transfer.
By replacing our 3D molecular mixture representation with SMILES, the overall performance drops. The 3D molecular structure conditioning is particularly important for system-level generalization as it better specifies the underlying chemical environment.
The peak constraint prevents spurious peak drift while encouraging physically plausible and amplitude changes, improving spectral fidelity. Channel-wise mean alignment enforces mass conservation by encouraging balanced dynamics with both positive and negative changes, reflected in the additional DILATE degradation when it is removed.

\noindent\textbf{Application: inferring plausible SEI formation pathways from predicted spectra.}
Beyond standard evaluation on held-out cells and unseen chemistries, \cref{fig:application} shows an application enabled by forward spectrum prediction. Given a predicted operando spectrum sequence, our model can be used to recover the most plausible SEI formation trajectory in an electrolyte system out of the training set, providing mechanistic interpretability rather than only waveform accuracy. Concretely, we decompose both the GT operando IR spectra and the spectrum sequence predicted by our model into different SEI-related contributions that reflect interphase growth and transformation, by adopting the same component disentanglement in~\cite{Leau2025Tracking}.
We apply this analysis to an unseen CoO/SuperP$\,|\,1$ M LiPF$_6$ in EC:DMC (1:1 v/v) system. The SEI components derived from our predictions closely track those from the GT, indicating that the model preserves chemically meaningful temporal signatures. The isolated SEI signals reproduce the literature-reported trend: carbonate products dominate, with an early component showing bands consistent with DMDOHC together with a semi-carbonate feature near $1640~\mathrm{cm^{-1}}$, followed by a later-emerging component assigned to LMC, including the characteristic band near $1355~\mathrm{cm^{-1}}$.

\vspace{-1mm}
\section{Conclusion}
\vspace{-0.5mm}
\jy{We introduce Operando IR Prediction, a novel task supported by OpIRSpec-7K, the first large-scale experimental dataset for spectral dynamics, and the OpIRBench benchmark. Our proposed model, ABCC, captures non-stationary electrode-electrolyte dynamics through Chemical Flow modeling, two-stream disentanglement, and physics-informed constraints. ABCC significantly outperforms static generation, video, and time-series baselines, offering a scalable computational alternative to costly operando experiments. 
}

\section*{Impact Statement}
This paper presents a framework designed to advance the intersection of Machine Learning and electrochemical science. By introducing the task of Operando IR Prediction, the newly constructed dataset and benchmark, and the ABCC framework, this work carries several potential societal and scientific impacts:

\noindent\textbf{Sustainability and Clean Energy.} Lithium-ion batteries (LIBs) are the foundation of the global transition toward sustainable energy and electromobility. By providing tools to better understand the Solid Electrolyte Interphase (SEI), this research contributes to improving battery performance, longevity, and safety.

\noindent\textbf{Democratization of Battery Research.} Currently, operando IR spectroscopy is an expensive and technically demanding diagnostic tool accessible only to a few elite laboratories. Our AI-driven approach offers a computational pathway to forecast spectral fingerprints, potentially allowing standard research facilities to bypass high experimental costs.

\noindent\textbf{Advancement of AI for Science (AI4Science).} This work introduces the first large-scale operando spectral dataset, OpIRSpec-7K, and a benchmark to scale AI-driven discovery in electrochemical dynamics. The inclusion of physics-informed constraints, such as mass conservation and peak drift regularization, ensures that generative models remain grounded in physical reality.

\bibliography{example_paper}
\bibliographystyle{icml2026}

\newpage
\appendix
\onecolumn




\section{Related Works on Physics-ML Hybrids}
\label{appendix:relatedwork-hybrids}
Physics-ML hybrids combine mechanistic models with data-driven learning to improve robustness and sample efficiency. Representative directions include physics-informed neural networks (PINNs), pseudo‑two‑dimensional (P2D) predictors, and differentiable parameter estimation. PINNs incorporate governing equations into the training objective so that predictions satisfy physical constraints under sparse or noisy supervision~\cite{Karniadakis2021Physics-informed,Shukla2022Scalable}. In batteries, P2D-informed neural predictors and other hybrid schemes leverage electrochemical models to provide synthetic training data, impose priors, or regularize latent state estimation for degradation and performance forecasting~\cite{Li2021Physics-informed,Tu2021Integrating,Singh2023Hybrid,Amiri2024Lithium-ion}. Related differentiable fitting frameworks, enabled by automatic differentiation, jointly optimize physical and neural components and have been adopted in areas such as electron microscopy and quantum chemistry~\cite{Malik2025Hybrid,Zubatiuk2021Development}.

However, our operando IR prediction poses additional challenges: reaction pathways, solvent reorganization, and SEI growth are only indirectly reflected in spectra and are difficult to parameterize with compact state variables, which complicates direct integration with standard hybrid formulations~\cite{Amiri2024Lithium-ion,Li2021Physics-informed,Tu2021Integrating,Singh2023Hybrid}. Moreover, many physics-ML hybrids are not designed to represent and propagate dynamics directly in spectral space~\cite{Karniadakis2021Physics-informed,Shukla2022Scalable,Zubatiuk2021Development,Malik2025Hybrid}. These gaps motivate an end-to-end framework that models voltage-conditioned temporal evolution while maintaining physical consistency. Our ABCC framework addresses this need through chemical flow modeling, mixture-level disentanglement, and physics-informed constraints tailored to operando spectral sequences.

\section{Related Works on Spectra Generation}
\label{appendix:relatedwork-spectra}
Substantial progress has been made in predicting molecular spectra, including IR, NMR, Raman, UV-Vis, and mass spectra, from molecular structure. Classical pipelines rely on quantum chemistry calculations such as DFT, which are accurate but computationally costly and often restricted to static settings and relatively small systems~\cite{Gastegger2017Machine,mcgill2021predicting,Pracht2024Efficient,Beckmann2022Infrared,Tsitsvero2023NMR,Hou2025Accurate}. To improve scalability, recent ML models, including message passing networks and graph transformers, enable high-throughput spectrum prediction from molecular graphs or SMILES, with multimodal and ensemble strategies improving robustness~\cite{mcgill2021predicting,Bhatia2025Leveraging}.

However, most existing approaches focus on static structure-to-spectrum mapping and typically omit time dependence, environmental factors, and process variables that drive spectral evolution~\cite{mcgill2021predicting,Pracht2024Efficient,Bhatia2025Leveraging,Beckmann2022Infrared,Tsitsvero2023NMR,Hou2025Accurate}. ML-accelerated molecular dynamics and anharmonic modeling can capture thermal effects, yet they still depend on extensive trajectory sampling and are not designed for operando forecasting~\cite{Bhatia2025Leveraging,Beckmann2022Infrared,Gastegger2017Machine,Tsitsvero2023NMR}. In particular, key controls for electrochemical operando IR, such as voltage histories and electrolyte composition, are rarely incorporated as explicit conditions.
These gaps motivate a dedicated formulation for operando IR prediction, where the goal is to forecast a spectrum sequence from a single initial measurement together with process conditions. This requires models that can represent sequential chemical transformations, mixture-level interactions, and physically plausible peak evolution under electrochemical driving.

\section{Details of OpIRBench and Evaluation Protocol}
\label{appendix:OpIRBench}
\noindent\textbf{Problem Definition.}
We formulate operando IR spectral prediction as a conditional generative modeling problem of an electrochemical experiment monitored by operando optical fiber IR spectroscopy. Given a voltage $v_m$ within the voltage sequence $V \in \mathbb{R}^{M}$ that drives SEI formation and electrolyte decomposition, the electrolyte as our defined mixture $\{X: R, …\}$ where $X\in{\mathbb{R}^{k\times3}}$ is 3D molecule structure and $R$ is its proportion, and the initial spectrum $\mathbf{s}_{v_0} \in \mathbb{R}^{N_f}$, our goal is to predict the IR spectrum $\mathbf{s}_{v_m} \in \mathbb{R}^{N_f}$, where $N_f$ denotes the number of wavenumber channels. The wavenumber range is truncated to $[f_{\min}, f_{\max}]$ (typically $[800, 2000]$ cm$^{-1}$) to focus on the chemically informative mid-IR region according to common practice~\cite{Leau2025Tracking}.

\noindent\textbf{Evaluation Metrics.}
We assess prediction quality with complementary metrics that reflect distinct requirements of operando IR modeling:

\textit{1)  Reconstruction errors} (\textbf{MAE}, \textbf{MSE}, \textbf{RMSE}) quantify pointwise fidelity between the predicted spectrum $\hat{\mathbf{s}}$ and the ground truth $\mathbf{s}$. 

\textit{2) Spectral similarity} targets shape consistency. \textbf{Spectral Angle Mapper} (SAM) measures the angle
$\theta=\arccos\!\left(\frac{\hat{\mathbf{s}}\cdot\mathbf{s}}{\|\hat{\mathbf{s}}\|\|\mathbf{s}\|}\right)$,
which is insensitive to global intensity scaling and therefore suited to operando drift. \textbf{Cosine similarity} provides a closely related normalized measure, and $\mathbf{R^2}$ \textbf{coefficient} summarizes how well variance across wavenumbers is explained.

\textit{3) Temporal coherence} evaluates whether predictions preserve physically plausible dynamics. Motivated by our RMS spectral-change analysis in~\cref{fig:data_analysis_combined}, where most frames exhibit small solvent-driven fluctuations interspersed with sparse jump-like SEI reaction events, we additionally include a \textbf{Dynamic Distortion Factor} (DILATE)~\cite{9721108} to emphasize accurate localization and magnitude of abrupt transitions while tolerating minor variations.

\noindent\textbf{Dataset Split.}
We establish our OpIRBench under two rigorous split protocols: Random Split to assess sequence prediction accuracy and System Split to evaluate generalization to unseen electrochemical systems. Under the Random Split, the dataset contains 6,407 training instances and 711 test instances. Under the System Split, the data are partitioned by battery electrochemical system, i.e., the electrolyte composition, where a system is defined by the working electrode, electrolyte recipe, and counter electrode. To enable evaluation of generalization across experimental configurations, OpIRBench adopts a system split protocol in which all spectra from a given system are assigned entirely to either training or testing, with no overlap in electrode or electrolyte conditions between splits. There are 8 systems used for training (5,796 instances) and 2 held-out unseen systems reserved for testing (1,322 instances).

Under this setting, the training split comprises eight systems spanning multiple electrodes and carbonate-based electrolytes. Specifically, it includes Cu$|$1\,M LiPF$_6$ in DMC$|$Li (693 data points), LTO$|$1\,M LiPF$_6$ in EC:DMC (1:1 w/w)$|$Li (661 points), Cu$|$1\,M LiPF$_6$ in EC:DMC (1:1 w/w)$|$Li (749 points), Sn$|$1\,M LiPF$_6$ in EC:DMC (1:1 w/w)$|$Li (767 points), and an intermittent LTO/SuperP (90/10) system with 1\,M LiPF$_6$ in EC:DMC (1:1 w/w)$|$Li (2,286 points). The training split further covers additive and solvent variations on Cu, including Cu$|$1\,M LiPF$_6$ in EC:DMC (1:1 w/w) + 2\% VC$|$Li (612 points), Cu$|$1\,M LiPF$_6$ in EMC$|$Li (14 points), and Cu$|$1\,M LiPF$_6$ in 0.3\,EC + 0.7\,EMC$|$Li (14 points).
The test split contains two held-out systems designed to probe transfer to new electrodes and electrolyte conditions: CoO/SuperP (90/10, SPEX 5 min in Eth/CMC, intermittent)$|$1\,M LiPF$_6$ in EC:DMC (1:1 w/w)$|$Li (1,308 data points), and Cu$|$1.2\,M LiPF$_6$ in EC$|$Li (14 points).
For machine readability, solvents are annotated with structure identifiers (e.g., DMC (dimethyl carbonate): \texttt{COC(=O)OC}, EMC (ethyl methyl carbonate): \texttt{CCOC(=O)OC}, EC (ethylene carbonate): \texttt{C1COC(=O)O1}, VC (vinylene carbonate): \texttt{C1=COC(=O)O1}).
We apply automated preprocessing commonly used in the community, including informative wavenumber truncation, scale clipping, and normalization to reduce global intensity drift.

\section{Implementation Details of ABCC}
\label{appendix:ABCC}
\textbf{Spectral-to-Image Conversion.} We render each 1D spectrum as a $256 \times 256$ grayscale image via Gaussian energy mapping. We use a margin $m=0.08$, base line thickness $\sigma_0=1.8$ pixels, and an amplitude-adaptive thickness
$\sigma = \sigma_0/3 + 1.8 \cdot |y-y_c|/y_c \cdot \sigma_0$,
where $y_c$ denotes the image centerline. Intensities are clipped to $\pm 0.4$ and re-scaled with contrast gain $g=180$.

\textbf{VAE.} We use the pretrained Stable Diffusion VAE \texttt{sd-vae-ft-ema} to encode images into $4 \times 32 \times 32$ latents.

\textbf{MeanFlow Architecture.} Our MeanFlow backbone is SiT-B/2 with hidden size 768, 12 transformer blocks, 12 attention heads, and patch size 2. The model is conditioned on the initial spectrum latent, electrochemical parameters (voltage), and electrolyte molecular embeddings.

\textbf{Explicit Residual Flow.} We add an explicit residual branch to model spectral variations using a separate variation image stream, with clip scale 0.1 and thickness 1.0.

\textbf{Training.} We train for 120 epochs with batch size 160 using AdamW and bf16 mixed precision on a machine with 2x NVIDIA A100 GPUs and 1x Intel Xeon(R) Gold 6258R processors. Flow matching adopts logit-normal time sampling with $\mu=-0.4$ and $\sigma=1.0$, bootstrap ratio 0.25, and adaptive loss weighting with $p=0.75$. We use classifier-free guidance with $\omega=1.0$ and $\kappa=0.5$.

\section{Baselines and Details}
\label{appendix:baselines}
To contextualize our study, we benchmark against representative methods from three related lines of research. First, we compare with state-of-the-art \textbf{IR spectrum generation} models, including NNMol-IR~\cite{ABDULAL2024141603} and MACE4IR~\cite{bhatia2025mace4irfoundationmodelmolecular}. These methods target static spectra and predict a single IR response for a given molecular structure. In contrast, our task models operando IR dynamics, where spectra evolve under voltage. To our knowledge, this work provides the first explicit formulation of voltage-conditioned IR spectral dynamics, which introduces temporal dependencies and control signals absent from prior IR generation methods. 
To establish strong baselines for high-capacity dynamic modeling, we adapt leading \textbf{video generation} models, CogVideoX~\cite{yangcogvideox} and Pyramid-Flow~\cite{jinpyramidal}, to treat IR trajectories as sequences with long-range correlations. Besides, we include a \textbf{time-series forecasting} baseline, STDN~\cite{cao2025spatiotemporal}, as a competitive model designed to capture structured temporal variation. Together, these baselines cover static IR generation, generic sequence generation, and dedicated forecasting. We also include random spectrums as a straightforward lower-bound baseline.

\textbf{Static IR Generators.} We include NNMol-IR and MACE4IR, which predict equilibrium IR spectra from molecular structure and do not model voltage-conditioned operando sequences. For NNMol-IR, SMILES strings are converted to a 3D conformer using RDKit, yielding atomic numbers together with 3D molecular coordinates (XYZ). The model then predicts a single, static IR spectrum for the given structure. To construct a spectrum sequence for operando evaluation, we repeat the same predicted spectrum across all time steps, since the method provides no mechanism to condition on voltage or current.
For MACE4IR, we generate equilibrium conformers of the molecular geometry in XYZ form, run short molecular dynamics, and compute IR spectra from dipole autocorrelation. We also repeat equilibrium spectra to adapt it to sequence prediction.

\textbf{Video Diffusion Baselines.} We adapt CogVideoX and Pyramid-Flow by replacing text prompts with the voltage profile and a cell-configuration embedding, conditioning with initial spectrum latent, then fine-tune on our operando dataset using the same training and evaluation splits. For CogVideoX, we finetune on its CogVideoX-5b-I2V pretrained weight and on latent to latent variant. For Pyramid-Flow, we finetune on its pyramid-flow-miniflux pretrained weight. For the video diffusion baselines, we report the best-performed intermediate checkpoints at training steps $\{2,4,20,60\}$.

\textbf{Time-Series Forecasting Baselines.} We adapt STDN by replacing its original temporal embedding, which encodes day of week and time of day, with the voltage profile. We also replace traffic node features with the initial spectrum $S_0$, which serves as the spatial conditioning input through the model’s spatial embedding module. Then the decoder generates the spectrum sequence using STDN’s trend and seasonality decomposition. We use the released pretrained STDN weights directly, without task-specific fine-tuning, to enable a zero-shot baseline.

\section{Future Works}
\label{appendix:future}
\jy{The introduction of OpIRSpec-7K and the ABCC framework establishes a foundation for predictive electrochemical spectroscopy. Future work will extend this framework to diverse battery chemistries and automated discovery pipelines.}
In future work, we will also explicitly model spectrum noise and uncertainty, enabling the framework to propagate confidence through the spectrum to waveform conversion and downstream prediction. We will also develop a more principled spectral rendering and inversion pipeline together with systematic diagnostics to quantify fidelity on sharp peaks and weak shoulder features. This direction will provide controllable, verifiable spectrum representations and improve robustness when operating conditions or instrumentation induce noise. Additionally, other formulations for diffusion models or flow matching models are also pausible for this task, and we leave this design space's further exploration to future work.

\end{document}